\documentclass[12pt,anonymous]{article}

\usepackage{sbc-template}
\usepackage{xcolor}
\usepackage{verbatim}
\usepackage{graphicx,url}
\usepackage[utf8]{inputenc}
\usepackage[english]{babel}
\usepackage{subcaption}
\usepackage{booktabs}
\usepackage{algorithm}
\usepackage{algorithmic}
\newcommand{\lista}{\mathcal{T}}

\title{Fast Frame Rate Estimation in Electromagnetic Side-Channel Attacks on Public Systems}

\author{
Alyson Isaluski\inst{1},
Leonardo Teodoro\inst{1},
Kleber V. Cardoso\inst{2},
Antonio Oliveira-Jr\inst{2, }\inst{3},\\
Saulo Queiroz\inst{1}
}
\address{
Federal University of Technology -- Paraná (UTFPR)\\
Ponta Grossa -- PR -- Brazil
\email{\{alysonisaluski, lteodoro\}@alunos.utfpr.edu.br}, \email{\{kleber, antoniojr\}@ufg.br},
\email{sauloqueiroz@utfpr.edu.br}
\nextinstitute
Federal University of Goiás (UFG)\\
Goiânia -- GO -- Brazil.
\nextinstitute
Fraunhofer Portugal AICOS, Porto 4200-135, Portugal.%
}

\begin{document}

\maketitle

\begin{abstract}
Frame refresh rate estimation is a fundamental step in identifying
compromising harmonic frequencies in electromagnetic side-channel attacks.
Methods based on discrete linear autocorrelation (DLA) are robust across
different scenarios and require a computational complexity of $O(N\log N)$
for a signal containing $N$ samples. This work proposes reducing this
complexity to $O(N)$ by exploiting prior knowledge of the target system's
display resolution, as is available for certain models of the Brazilian
electronic voting machine. Experiments using software-defined radios show
that the proposed method preserves the accuracy of the conventional
approach in the evaluated scenarios.
\end{abstract}

\section{Introduction}
Electromagnetic side-channel attacks (e-SCAs) refer to a class of security
threats in which sensitive information is inferred from unintentional
electromagnetic emissions produced by the movement of electric charges.
These attacks are commonly known as TEMPEST
attacks~\cite{tempestclass-ieeeaccess2025}. In this work, we use the term
TEMPEST specifically to denote e-SCAs targeting video devices.

A critical challenge in TEMPEST attacks is estimating the frame refresh
rate (FRR) from the leaked signal, which is necessary to reconstruct the
intercepted image and helps identify a compromising harmonic frequency.
Various methods for FRR estimation have been considered in the
literature~\cite{deepl-electcompat-2025},~\cite{deepl-larroca-2024},
~\cite{freqdomainfrr-2010}. Among these, discrete linear autocorrelation
(DLA) stands out as a popular method in practical TEMPEST libraries
because of its effective estimation across diverse
scenarios~\cite{grtempest-2022},~\cite{tempestsdr-marinov}.

This work proposes a method to reduce the computational complexity of
DLA from $O(N\log N)$ to $O(N)$ in scenarios where the target system's
display resolution is known beforehand, as observed
in~\cite{publicTEMPEST2026} for some Brazilian electronic voting machine
(UEB) models currently in use. The proposed method is validated using
real TEMPEST signals obtained in experiments with software-defined
radio (SDR).

\section{TEMPEST Signal Model}
This section introduces the theoretical foundations of this work.
First, Subsection~\ref{subsec:background} presents the TEMPEST signal
parameters required by the proposed method. Next,
Subsection~\ref{subsec:dla} presents the DLA method, whose computational
complexity is optimized in this work.

\subsection{Basic TEMPEST Parameters}\label{subsec:background}
The VGA signal assumed in this work can be modeled as a sum of
rectangular pulses:
\begin{equation}
x(t) = \sum_{n=-\infty}^{\infty} x[n]p(t-n T_p),
\end{equation}
where $x(t)$ represents the real-valued continuous-time signal at time
$t$. The period (i.e., duration) $T_p$ of each pixel (pulse) in the cable
depends on the display resolution and frame refresh rate.
$T_p$ can be defined as
\begin{equation}
T_p = \frac{1}{P_x P_y f_v} \quad \textrm{s},
\end{equation}
where $P_x$, $P_y$, and $f_v$ denote the number of pixels per line
(including synchronization pixels, i.e., \emph{blanking} pixels), the
number of lines per display frame, and the refresh rate, respectively.
TEMPEST attacks are made possible by distortions inherent in the analog
components of the video system, which cause leakage at harmonic
frequencies that are multiples of the pixel
rate~\cite{tempestclass-ieeeaccess2025},~\cite{lvds-ieeetifs-25},~\cite{disp-eleccompa-2019}.
An adversary who tunes an SDR to one of these frequencies can recover
the image carried by the signal.

\subsection{FRR Estimation Using the DLA Method}\label{subsec:dla}
After acquiring the raw signal using an SDR in a TEMPEST attack, the
frame refresh rate $f_v$ must be estimated. The DLA method relies on
identifying the dominant peak of the autocorrelation function. For a
discrete complex-valued signal $x[n]$ containing $N$ samples, the
autocorrelation at \emph{lag} $\tau$ is given by
\begin{eqnarray}
R_{xx}(\tau) &=& \sum_{n=\tau}^{N-1}
x[n]x^{*}[n-\tau],
\qquad
0 \leq \tau < N, \label{eqn:dla}
\end{eqnarray}
where $x^{*}[n]$ denotes the complex conjugate of $x[n]$. For a sampling
rate $f_s$, the FRR estimate is obtained as
\begin{eqnarray}
f_v &\approx& f_s/\tau_{\max}, \label{eqn:fv}
\end{eqnarray}
where $\tau_{\max}$ is the \emph{lag} associated with the highest peak in
(\ref{eqn:dla}).

A direct implementation of DLA evaluates the autocorrelation at all
\emph{lags} of interest, resulting in a complexity of $O(N^2)$. To reduce
this cost, widely used TEMPEST
tools~\cite{grtempest-2022},~\cite{tempestsdr-marinov} rely on the
Wiener--Khinchin theorem, according to which autocorrelation can be
obtained from the inverse Fourier transform of the signal's power
spectrum. Thus, using the fast Fourier transform (FFT) algorithm and its
inverse (IFFT), the DLA method runs in $O(N\log N)$ time. Although
efficient in the general case, the computational cost of processing
signals containing millions of samples---such as the TEMPEST signals
considered in this work---may become a limiting factor in critical
applications subject to stringent timing constraints, as demonstrated
in~\cite{fastenough-2022}.

\section{The \emph{Fast DLA} (F-DLA) Method}\label{sec:fdla}
The proposed method, called F-DLA (\emph{fast} DLA), reduces the
computational complexity of conventional DLA by exploiting TEMPEST
scenarios in which the parameters $P_x$ and $P_y$ may be known
\emph{a priori}. Although this is a strong assumption in general TEMPEST
scenarios, the assumption that the target system's parameters are
unknown can be relaxed in contexts governed by transparency principles,
as in the Brazilian electoral system. Public specifications establish
resolutions of $1280\times768$ for the UE2013/UE2015
models~\cite{tresc_res_ue2015_ue2015} and at least $1280\times720$ for the
UE2020 model~\cite{tse_ue2020_security_specs_resolucao}. Thus, even without
prior knowledge of the model in use, the VESA standard restricts the set
of candidate FRRs to only $R=4$ values for these resolutions: 60, 75, 85,
and 120 Hz~\cite{vesa_dmt_2013}. Accordingly, the first significant
autocorrelation peak is expected at one of the \emph{lags}
$\{f_s/60,f_s/75,f_s/85,f_s/120\}$, with $R=O(1)$ with respect to $N$.

Correct operation of the DLA method assumes that channel-induced
distortions do not shift the dominant autocorrelation peak beyond the
interval associated with the correct FRR. Otherwise, even conventional
DLA would produce an incorrect estimate. Therefore, in scenarios where
conventional DLA operates correctly, the shift $\delta$ must remain
bounded by a constant independent of $N$. Consequently, only the
\emph{lags} within a neighborhood of $\delta$ samples around the $R$
candidates need to be evaluated. Since $R=O(1)$ and $\delta=O(1)$, the set
$\lista$ of candidate \emph{lags} whose autocorrelation values must be
evaluated has size
\begin{eqnarray}
   |\lista|=(2\delta + 1)R = O(1) \label{eqn:listat}.
\end{eqnarray}

Similarly to Eq.~(\ref{eqn:fv}), F-DLA estimates the FRR as
\begin{eqnarray}
f_v &\approx& \frac{f_s}{\tau_{F\textrm{-}DLA}} \label{eqn:fvfdla},
\end{eqnarray}
where $\tau_{F\textrm{-}DLA}\in\lista$ is the \emph{lag} that maximizes the
autocorrelation function:
\begin{eqnarray}
  R_{xx}(\tau_{F\textrm{-}DLA})=\max_{ \tau\in \lista } R_{xx}(\tau) \label{eqn:fdla}
\end{eqnarray}

Since each \emph{lag} in $\lista$ is evaluated in $O(N)$ time
(\ref{eqn:dla}) and this list has constant size (\ref{eqn:listat}), the
overall complexity of F-DLA is
\begin{eqnarray}
  T_{F-DLA}(N)=|\lista|N=O(N) \label{eqn:complexidade},
\end{eqnarray}
which is therefore lower than the $O(N\log N)$ complexity of conventional
DLA. Under severe channel degradation, the assumption $\delta=O(1)$ may
no longer hold, compromising FRR estimation by both DLA and F-DLA. Since
this work aims to validate F-DLA as an alternative to DLA in at least one
real-world scenario, evaluation under degraded channel conditions is
left for future work.

\section{Preliminary Results}\label{sec:analise}
This section compares the proposed F-DLA method with conventional DLA in
terms of execution time and the accuracy of the $f_v$ estimate, using
signals obtained in a real SDR-based TEMPEST experiment. The experiments
used a VGA monitor displaying the UEB interface (available
in~\cite{tse_voting_simulator}), configured with an FRR of $f_v=60$ Hz and
a display resolution of $1280\times 720$, compatible with the UE2020
model. For this configuration, $P_x=1650$ pixels and $P_y=750$ lines per
frame, according to the VESA standard~\cite{vesa_dmt_2013}. For further
details on the results presented in this section, as well as additional
results omitted owing to space limitations, the reader is referred to
the lead author's
repository\footnote{\url{https://github.com/AlysonIsa/SBSeg-FDLA}.}.

The video TEMPEST signal was captured using an Ettus USRP B200 SDR
connected to a digital TV antenna positioned $\approx$~20 cm from the
monitor, with the SDR operating at a sampling rate of $f_s=54$ MHz.
Both methods used $N=2^{21}$ samples. This number is sufficient to capture
the minimum of two consecutive frames, corresponding to
$2(f_s/f_v)=1.8\times10^6$ samples in the scenario considered. The
\texttt{xcorr} function from GNU Octave's \emph{signal} package was used
as the reference implementation of conventional DLA. For the proposed
F-DLA method, the smallest value of $\delta$ that produced the correct
$f_v$ estimate in the experiments was 1. Although preliminary, the
experimental results support the theoretical gains discussed in
Section~\ref{sec:fdla}. A more comprehensive experimental evaluation is
planned for an extended version of this work.

\begin{figure}[t]
    \centering
    \begin{subfigure}{0.46\linewidth}
        \centering
        \includegraphics[width=.95\linewidth]{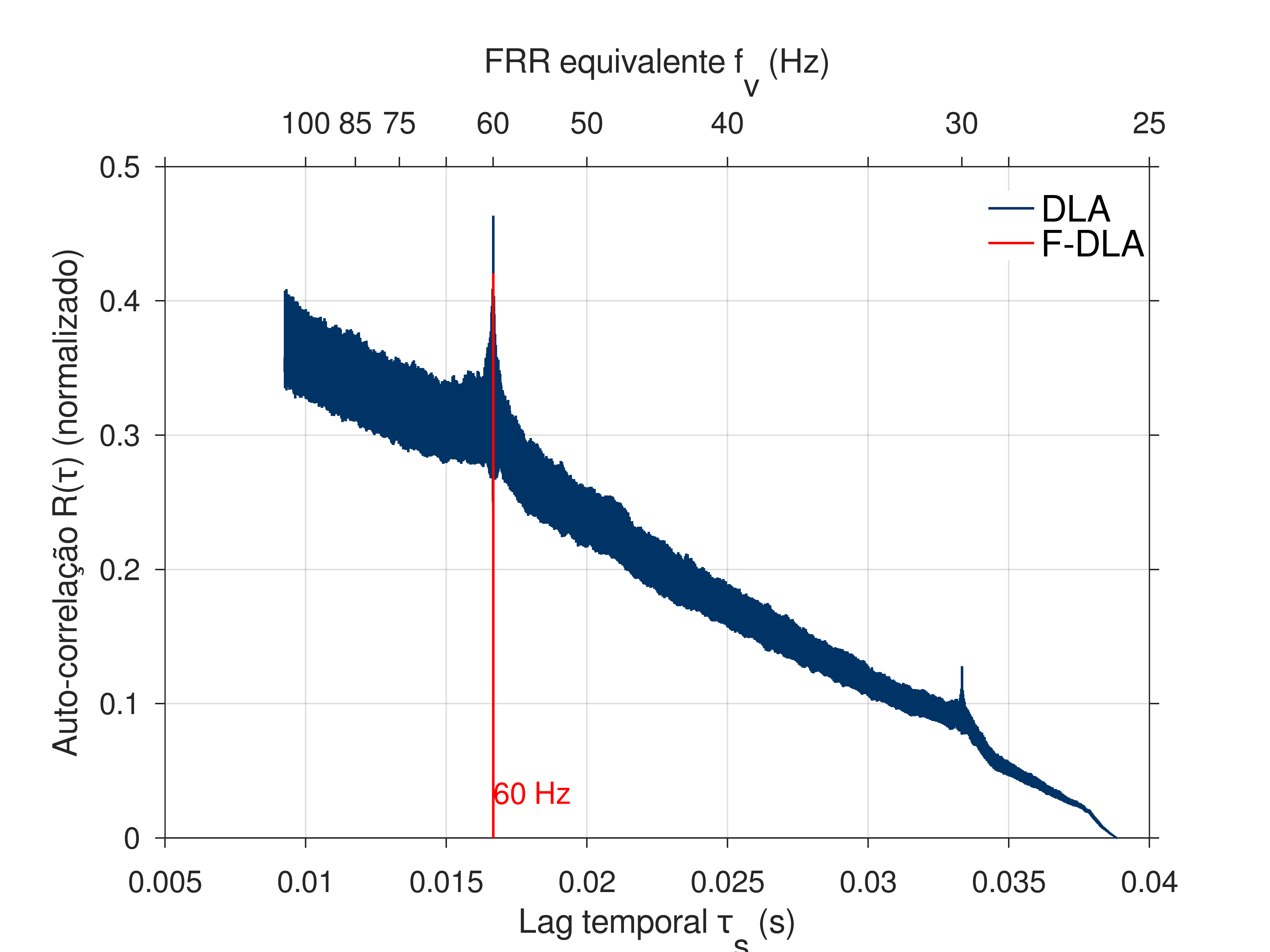}
        \caption{Autocorrelation function plot.}
        \label{fig:autocor}
    \end{subfigure}
    \hfill
    \begin{subfigure}{0.46\linewidth}
        \centering
        \includegraphics[width=0.74\linewidth]{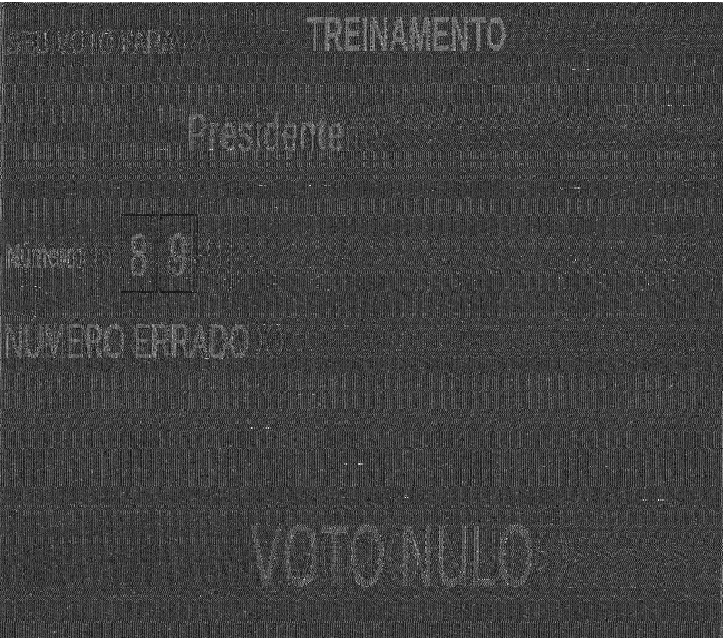}
        \caption{TEMPEST image reconstructed from the experiments.}
        \label{fig:imagem}
    \end{subfigure}
\caption{F-DLA (proposed) and DLA (conventional): FRR identification in an experimental TEMPEST channel. The autocorrelation peak at $\tau_s\approx 0.016667$ s obtained by both methods in (a) corresponds to the correct FRR of 60 Hz for the reconstructed image in (b).
}
    \label{fig:experimento}
\end{figure}

Figure~\ref{fig:autocor} shows the autocorrelation function of the TEMPEST
signal whose image, reconstructed using the gr-tempest
module~\cite{grtempest-2022}, is shown in Fig.~\ref{fig:imagem}. The time
\emph{lag} $\tau_s$, in seconds, was obtained using $\tau_s=\tau/f_s$.
The autocorrelation peak identified by F-DLA (red bar) at
$\tau_s=0.016667$ s coincides with that obtained by DLA (blue curve) and
corresponds to the expected FRR of $1/\tau_s\approx60$ Hz. However, F-DLA
produced this estimate in approximately 0.30 s, with a confidence
interval (CI) of $\pm0.01$ s (95\% confidence), whereas conventional DLA
required approximately 0.67 s, with a CI of $\pm0.02$ s. The observed
difference reflects implementation constants and optimizations in the
\texttt{xcorr} function, as well as the asymptotic orders of the
algorithms. This result indicates that restricting the set of candidate
\emph{lags} based on prior knowledge of $P_x$ and $P_y$, as in the case
of the UEB, can reduce the computational cost of estimation without
compromising FRR accuracy in the evaluated scenario.

\section{Conclusion}
This work presented \emph{Fast DLA} (F-DLA), a low-complexity alternative
for frame refresh rate estimation in TEMPEST attacks. The method
exploits scenarios where the target graphics system's parameters may be
known \emph{a priori}, restricting the set of candidate \emph{lags}
evaluated through autocorrelation. Under this assumption, the
computational complexity was shown to decrease from $O(N\log N)$ to
$O(N)$. Results from a real SDR-based TEMPEST experiment showed that
F-DLA preserves the FRR estimation accuracy of conventional DLA while
reducing processing time. Although preliminary, these results support
the theoretical analysis and indicate the approach's potential for
applications with more stringent computational constraints. Future work
will extend the experimental evaluation to different scenarios and
parameters and consider complexity optimization techniques based on
sparse FFTs, e.g.,~\cite{sic-magazine-2025}.

\section*{Acknowledgments}
This work was partially funded by the \emph{Advanced Multimodal Sensing}
(AIMS) project, supported by the \emph{Advanced Knowledge Center in
Immersive Technologies} (AKCIT), with funding from MCTI's PPI IoT program
under Agreement No.~057/2023 with EMBRAPII. The authors also thank the
Goiás Research Foundation (FAPEG) for the financial support provided for
this research (Project No.~64448878/2024).

\bibliographystyle{sbc}
\bibliography{refs}

\end{document}